\documentclass{article}
\usepackage{latexsym,alife9,pstricks}
\title{The influence of parsimony and randomness on complexity growth
  in Tierra}
\author{Russell K. Standish\\School Mathematics, UNSW, 2052
  Australia\\
http://parallel.hpc.unsw.edu.au/rks, r.standish@unsw.edu.au}

\def\citeyear(#1)#2{(#1)\nocite{#2}}

\begin{document}
\maketitle
\begin{abstract}
The issue of how to create open-ended evolution in an artificial
system is one the open problems in artificial life. This paper
examines two of the factors that have some bearing on this issue,
using the Tierra artificial life system. 

{\em Parsimony pressure} is a tendency to penalise more complex
organisms by the extra cost needed to reproduce longer genotypes,
encouraging simplification to happen. In Tierra, parsimony is
controlled by the \verb+SlicePow+ parameter. When full parsimony is
selected, evolution optimises the ancestral organism to produce
extremely simple organisms. With parsimony completely relaxed,
organisms grow larger, but not more complex. They fill up with
``junk''. This paper looks at scanning a range of \verb+SlicePow+ from
0.9 to 1 to see if there is an optimal value for generating
complexity.

Tierra (along with most ALife systems) use pseudo random number
generators. Algorithms can never create information, only destroy
it. So the total complexity of the Tierra system is bounded by the
initial complexity, implying that the individual organism complexity
is bounded. Biological systems, however, have plenty of sources of
randomness, ultimately dependent on quantum randomness, so do not have
this complexity limit. Sources of real random numbers exist for
computers called {\em entropy gatherers} --- this paper reports on the
effect of changing Tierra's pseudo random number generator for an
entropy gatherer.
\end{abstract}

\section{Introduction}

The issue of how to create open-ended evolution in an artificial
system is one the open problems in artificial life. This paper
examines two of the factors that have some bearing on this issue,
using the Tierra artificial life system\cite{Ray91}\footnote{Available
  from http://www.his.atr.jp/\~{}ray/tierra/}. Tierra is well known
artificial system, and well described in the literature, so only brief
details will be given here. The digital organisms in Tierra consist of
self-replicating codes written in a specially designed machine
language. The Tierra environment is a virtual operating system
executing the organism's code in a time shared manner. Evolution
proceeds through deliberately introduced mutations, copying errors and
instruction flaws. Organisms compete for CPU time and memory space
(called {\em soup}).

{\em Parsimony pressure} is a tendency to penalise more complex
organisms by the extra cost needed to reproduce longer genotypes,
encouraging simplification to happen. In Ray's earliest experiments
with Tierra, CPU time was allocated evenly between organisms,
favouring organisms with the shortest genomes. The time sharing system
was changed so that CPU time was allocated proportional to
$\ell^\mathtt{SlicePow}$. When \verb+SlicePow+=0, we have the original
maximal parsimony pressure. When \verb+SlicePow+=1, parsimony pressure
is removed entirely. In this case, organism length rapidly increases,
until the soup consists of one organism whose length is greater than
half of Tierra's memory. At this point, it can no longer reproduce,
and the soup dies (simulation stops).

But do organisms get more complex? For this purpose, we define
complexity to be the {\em algorithmic information} \cite{Li-Vitanyi97}
of the organism. Adami\citeyear(1998){Adami98a} introduced this
measure in an artificial life setting, and I \cite{Standish99a}
developed a technique for measuring this in the Tierra setting. In
\cite{Standish03a}, I report the first detailed study of a Tierra run.
Whilst organisms get longer, their complexity shows no sign of
increase at all. Their length comes from adding ``junk'' into their
genomes.

Obvious neither extremes of \verb+SlicePow+ leads to complexity
growth, but what if we tuned the parsimony pressure to modest values?
In previous experiments, I knew that \verb+SlicePow+$<0.9$ led to
shorter genomes, not longer, so in this paper, I scan a range of
\verb+SlicePow+ from 0.9 to 1 to see if there is an optimal value for
generating complexity.

Tierra (along with most ALife systems) use pseudo random number
generators. Pseudo random number generators are short algorithms
satisfying certain statistical tests for uniformity and independence.
However, being the product of an algorithm, the output of a pseudo
random number generator is not random by definition
\cite{Li-Vitanyi97}.  Algorithms can never create information, only
destroy it. The complexity of any sequence of numbers is closely
related to the length of the shortest algorithm that produces it, so
the total complexity of the Tierra system with pseudo random number
generators is bounded by its initial complexity, implying that the
individual organism complexity is bounded. Biological systems,
however, have plenty of sources of randomness, ultimately dependent on
quantum randomness, so do not have this complexity limit. 

The only way an algorithm can generate unbounded complexity is if it
is coupled to a source of real randomness --- a {\em random oracle}.
Random oracles feature in Douglas Adams's description of the {\em
  infinite improbability drive}: 
\begin{quote}
{\em a Bambleweeny 57 Sub-meson Brain
coupled to an atomic vector plotter suspended in a strong Brownian
Motion producer (say a nice hot cup of tea)} \cite[Chapt.
10]{Adams79}.
\end{quote}
 It turns out to be simple enough to create random
oracles: Geiger counters attached to radioactive sources
\cite{Gude85}\footnote{http://www.fourmilab.ch/hotbits/} and Lava
lamps\footnote{http://www.lavarnd.org/} are available through the
internet to provide sources of genuine randomness, however these
sources are limited to about 30 bytes per second.

Computers themselves have many different sources of random data
available. They often interact with the external environment (eg users
using keyboards, mice etc), and there is a small amount of randomness
in timings in the hard disk \cite{Jakobsson-etal98}. Programs that
harvest these sources of physical randomness are called {\em entropy
  gatherers}. Since unpredictability is important for cryptographic
applications, practical true random number generation has experienced
a lot of development in recent years. For example, the Linux operating
system includes an entropy gatherer in its kernel, available as
a character device at \verb+/dev/random+.

Unfortunately, entropy gatherers, like the internet available random
oracles, tend to be slow producers of randomness. HAVEGE
\cite{Seznec-Sendrier03}
\footnote{http://www.irisa.fr/caps/projects/hipsor/HAVEGE.html}
exploits many different sources of entropy with a modern computing
system using hand crafted assembly language routines to increase the
rate of entropy production by 3-4 orders of magnitude over the
techniques available in \verb+/dev/random+. This random stream is then
used to seed a lookup table accessed by a pseudo random number
generator to produce random numbers at similar rates to traditional
pseudo random number generators. The entropy of the resulting sequence
is less than a truly random sequence, but considerably higher than a
pseudo random generator.

In this paper, we replace the pseudo random number generator in Tierra
by calls to the HAVEGE generator, and compare what difference this
makes to growth of complexity.

\section{Measurement of Complexity in Tierra}

The most general definition of complexity of an object involves two levels of
description, a {\em micro-}description which is its implementation,
and a {\em macro-}description which is the abstract {\em meaning} of
an object. More than one microdescription can correspond to the same
macrodescription. If $\omega(\ell,x)$ is the number of
microdescriptions of length $\ell$ corresponding to macrodescription
$x$, then the complexity of $x$ is given by \cite{Standish01a}:
\begin{equation}\label{Complexity}
C(x) = \lim_{\ell\rightarrow\infty} \ell \log N - \log \omega(\ell,x).
\end{equation}
where $N$ is the size of the alphabet used for the microdescription.
Eq (\ref{Complexity}) converges extremely rapidly for $\ell >
C(x)/\log N$.
The base of the logarithm determines what units you are measuring
complexity in --- if it is base 2, then the units are bits. For
convenience, in this paper we will use base 32, corresponding to the
alphabet size of the Tierra instruction set. Complexity is
then measured in {\em instructions}. In order to measure the
complexity of an organism in Tierra, we simply need to count up the
number $\omega(\ell,x)$ of computer programs of length $\ell$ that are
equivalent to a given digital organism $x$.

Not so simple! The first problem is how to determine if two computer
programs are equivalent. The technique we use \cite{Standish03a}, is
to record the results of a tournament where an organism is pitted
pairwise against all genotypes recorded from a given Tierra run. Since
this includes the context that these organisms experienced, any
difference between two organism is expected to show up as a difference
in the results of the two tournaments.

The second problem is that the number of programs of length $\ell$ is
$32^\ell$, a computationally infeasible number. In \cite{Standish03a},
I show that an alternative measure $C_\mathrm{ss}$ is a good first
order estimate of the organismal complexity:
\begin{equation}\label{sse}
C_\mathrm{ss}=\ell-\sum_{i=1}\log_{32}n_i
\end{equation}
where $n_i$ is the number of mutations at site $i$ on the genome that
lead to differing phenotypes. 

This quantity is now very tractable, with a complete analysis of a
$10^{10}$ timestep Tierra run taking only a few hours on ac3's Linux
cluster --- comparable to the time taken to perform the original
Tierra run.\footnote{The analysis code is available from
  http://parallel.hpc.unsw.edu.au/getaegisdist.cgi/getdeltas/eco-tierra.3}

\section{Results}

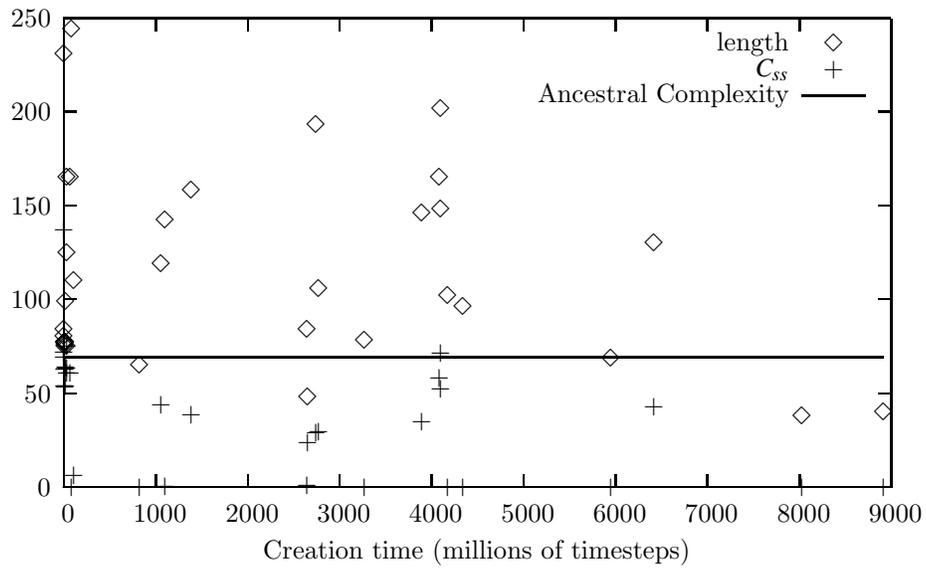
\begin{figure*}
\begin{center}
% GNUPLOT: LaTeX picture
\setlength{\unitlength}{0.240900pt}
\ifx\plotpoint\undefined\newsavebox{\plotpoint}\fi
\sbox{\plotpoint}{\rule[-0.200pt]{0.400pt}{0.400pt}}%
\begin{picture}(1500,900)(0,0)
\font\gnuplot=cmr10 at 10pt
\gnuplot
\sbox{\plotpoint}{\rule[-0.200pt]{0.400pt}{0.400pt}}%
\put(140.0,123.0){\rule[-0.200pt]{4.818pt}{0.400pt}}
\put(120,123){\makebox(0,0)[r]{ 0}}
\put(1419.0,123.0){\rule[-0.200pt]{4.818pt}{0.400pt}}
\put(140.0,270.0){\rule[-0.200pt]{4.818pt}{0.400pt}}
\put(120,270){\makebox(0,0)[r]{ 50}}
\put(1419.0,270.0){\rule[-0.200pt]{4.818pt}{0.400pt}}
\put(140.0,418.0){\rule[-0.200pt]{4.818pt}{0.400pt}}
\put(120,418){\makebox(0,0)[r]{ 100}}
\put(1419.0,418.0){\rule[-0.200pt]{4.818pt}{0.400pt}}
\put(140.0,565.0){\rule[-0.200pt]{4.818pt}{0.400pt}}
\put(120,565){\makebox(0,0)[r]{ 150}}
\put(1419.0,565.0){\rule[-0.200pt]{4.818pt}{0.400pt}}
\put(140.0,713.0){\rule[-0.200pt]{4.818pt}{0.400pt}}
\put(120,713){\makebox(0,0)[r]{ 200}}
\put(1419.0,713.0){\rule[-0.200pt]{4.818pt}{0.400pt}}
\put(140.0,860.0){\rule[-0.200pt]{4.818pt}{0.400pt}}
\put(120,860){\makebox(0,0)[r]{ 250}}
\put(1419.0,860.0){\rule[-0.200pt]{4.818pt}{0.400pt}}
\put(140.0,123.0){\rule[-0.200pt]{0.400pt}{4.818pt}}
\put(140,82){\makebox(0,0){ 0}}
\put(140.0,840.0){\rule[-0.200pt]{0.400pt}{4.818pt}}
\put(284.0,123.0){\rule[-0.200pt]{0.400pt}{4.818pt}}
\put(284,82){\makebox(0,0){ 1000}}
\put(284.0,840.0){\rule[-0.200pt]{0.400pt}{4.818pt}}
\put(429.0,123.0){\rule[-0.200pt]{0.400pt}{4.818pt}}
\put(429,82){\makebox(0,0){ 2000}}
\put(429.0,840.0){\rule[-0.200pt]{0.400pt}{4.818pt}}
\put(573.0,123.0){\rule[-0.200pt]{0.400pt}{4.818pt}}
\put(573,82){\makebox(0,0){ 3000}}
\put(573.0,840.0){\rule[-0.200pt]{0.400pt}{4.818pt}}
\put(717.0,123.0){\rule[-0.200pt]{0.400pt}{4.818pt}}
\put(717,82){\makebox(0,0){ 4000}}
\put(717.0,840.0){\rule[-0.200pt]{0.400pt}{4.818pt}}
\put(862.0,123.0){\rule[-0.200pt]{0.400pt}{4.818pt}}
\put(862,82){\makebox(0,0){ 5000}}
\put(862.0,840.0){\rule[-0.200pt]{0.400pt}{4.818pt}}
\put(1006.0,123.0){\rule[-0.200pt]{0.400pt}{4.818pt}}
\put(1006,82){\makebox(0,0){ 6000}}
\put(1006.0,840.0){\rule[-0.200pt]{0.400pt}{4.818pt}}
\put(1150.0,123.0){\rule[-0.200pt]{0.400pt}{4.818pt}}
\put(1150,82){\makebox(0,0){ 7000}}
\put(1150.0,840.0){\rule[-0.200pt]{0.400pt}{4.818pt}}
\put(1295.0,123.0){\rule[-0.200pt]{0.400pt}{4.818pt}}
\put(1295,82){\makebox(0,0){ 8000}}
\put(1295.0,840.0){\rule[-0.200pt]{0.400pt}{4.818pt}}
\put(1439.0,123.0){\rule[-0.200pt]{0.400pt}{4.818pt}}
\put(1439,82){\makebox(0,0){ 9000}}
\put(1439.0,840.0){\rule[-0.200pt]{0.400pt}{4.818pt}}
\put(140.0,123.0){\rule[-0.200pt]{312.929pt}{0.400pt}}
\put(1439.0,123.0){\rule[-0.200pt]{0.400pt}{177.543pt}}
\put(140.0,860.0){\rule[-0.200pt]{312.929pt}{0.400pt}}
\put(789,21){\makebox(0,0){Creation time (millions of timesteps)}}
\put(140.0,123.0){\rule[-0.200pt]{0.400pt}{177.543pt}}
\put(1279,820){\makebox(0,0)[r]{length}}
\put(1299,235){\raisebox{-.8pt}{\makebox(0,0){$\Diamond$}}}
\put(1427,241){\raisebox{-.8pt}{\makebox(0,0){$\Diamond$}}}
\put(523,265){\raisebox{-.8pt}{\makebox(0,0){$\Diamond$}}}
\put(259,315){\raisebox{-.8pt}{\makebox(0,0){$\Diamond$}}}
\put(999,326){\raisebox{-.8pt}{\makebox(0,0){$\Diamond$}}}
\put(142,344){\raisebox{-.8pt}{\makebox(0,0){$\Diamond$}}}
\put(145,344){\raisebox{-.8pt}{\makebox(0,0){$\Diamond$}}}
\put(142,347){\raisebox{-.8pt}{\makebox(0,0){$\Diamond$}}}
\put(141,350){\raisebox{-.8pt}{\makebox(0,0){$\Diamond$}}}
\put(142,350){\raisebox{-.8pt}{\makebox(0,0){$\Diamond$}}}
\put(142,350){\raisebox{-.8pt}{\makebox(0,0){$\Diamond$}}}
\put(142,350){\raisebox{-.8pt}{\makebox(0,0){$\Diamond$}}}
\put(143,350){\raisebox{-.8pt}{\makebox(0,0){$\Diamond$}}}
\put(612,353){\raisebox{-.8pt}{\makebox(0,0){$\Diamond$}}}
\put(140,359){\raisebox{-.8pt}{\makebox(0,0){$\Diamond$}}}
\put(140,371){\raisebox{-.8pt}{\makebox(0,0){$\Diamond$}}}
\put(522,371){\raisebox{-.8pt}{\makebox(0,0){$\Diamond$}}}
\put(767,406){\raisebox{-.8pt}{\makebox(0,0){$\Diamond$}}}
\put(143,415){\raisebox{-.8pt}{\makebox(0,0){$\Diamond$}}}
\put(743,424){\raisebox{-.8pt}{\makebox(0,0){$\Diamond$}}}
\put(540,435){\raisebox{-.8pt}{\makebox(0,0){$\Diamond$}}}
\put(156,447){\raisebox{-.8pt}{\makebox(0,0){$\Diamond$}}}
\put(293,474){\raisebox{-.8pt}{\makebox(0,0){$\Diamond$}}}
\put(145,491){\raisebox{-.8pt}{\makebox(0,0){$\Diamond$}}}
\put(1067,506){\raisebox{-.8pt}{\makebox(0,0){$\Diamond$}}}
\put(299,542){\raisebox{-.8pt}{\makebox(0,0){$\Diamond$}}}
\put(702,553){\raisebox{-.8pt}{\makebox(0,0){$\Diamond$}}}
\put(732,559){\raisebox{-.8pt}{\makebox(0,0){$\Diamond$}}}
\put(340,589){\raisebox{-.8pt}{\makebox(0,0){$\Diamond$}}}
\put(145,609){\raisebox{-.8pt}{\makebox(0,0){$\Diamond$}}}
\put(150,609){\raisebox{-.8pt}{\makebox(0,0){$\Diamond$}}}
\put(730,609){\raisebox{-.8pt}{\makebox(0,0){$\Diamond$}}}
\put(536,692){\raisebox{-.8pt}{\makebox(0,0){$\Diamond$}}}
\put(732,718){\raisebox{-.8pt}{\makebox(0,0){$\Diamond$}}}
\put(140,804){\raisebox{-.8pt}{\makebox(0,0){$\Diamond$}}}
\put(152,842){\raisebox{-.8pt}{\makebox(0,0){$\Diamond$}}}
\put(1349,820){\raisebox{-.8pt}{\makebox(0,0){$\Diamond$}}}
\put(1279,779){\makebox(0,0)[r]{$C_{ss}$}}
\put(1299,123){\makebox(0,0){$+$}}
\put(1427,123){\makebox(0,0){$+$}}
\put(523,193){\makebox(0,0){$+$}}
\put(259,123){\makebox(0,0){$+$}}
\put(999,123){\makebox(0,0){$+$}}
\put(142,281){\makebox(0,0){$+$}}
\put(145,303){\makebox(0,0){$+$}}
\put(142,308){\makebox(0,0){$+$}}
\put(141,282){\makebox(0,0){$+$}}
\put(142,309){\makebox(0,0){$+$}}
\put(142,309){\makebox(0,0){$+$}}
\put(142,309){\makebox(0,0){$+$}}
\put(143,282){\makebox(0,0){$+$}}
\put(612,123){\makebox(0,0){$+$}}
\put(140,327){\makebox(0,0){$+$}}
\put(140,336){\makebox(0,0){$+$}}
\put(522,126){\makebox(0,0){$+$}}
\put(767,123){\makebox(0,0){$+$}}
\put(143,312){\makebox(0,0){$+$}}
\put(743,123){\makebox(0,0){$+$}}
\put(540,210){\makebox(0,0){$+$}}
\put(156,141){\makebox(0,0){$+$}}
\put(293,252){\makebox(0,0){$+$}}
\put(145,311){\makebox(0,0){$+$}}
\put(1067,249){\makebox(0,0){$+$}}
\put(299,124){\makebox(0,0){$+$}}
\put(702,226){\makebox(0,0){$+$}}
\put(732,277){\makebox(0,0){$+$}}
\put(340,237){\makebox(0,0){$+$}}
\put(145,344){\makebox(0,0){$+$}}
\put(150,302){\makebox(0,0){$+$}}
\put(730,294){\makebox(0,0){$+$}}
\put(536,209){\makebox(0,0){$+$}}
\put(732,333){\makebox(0,0){$+$}}
\put(140,527){\makebox(0,0){$+$}}
\put(152,123){\makebox(0,0){$+$}}
\put(1349,779){\makebox(0,0){$+$}}
\sbox{\plotpoint}{\rule[-0.400pt]{0.800pt}{0.800pt}}%
\put(1279,738){\makebox(0,0)[r]{Ancestral Complexity}}
\put(1299.0,738.0){\rule[-0.400pt]{24.090pt}{0.800pt}}
\put(140,327){\usebox{\plotpoint}}
\put(140.0,327.0){\rule[-0.400pt]{310.038pt}{0.800pt}}
\end{picture}
\end{center}
\caption{Plot of length and $C_ss$ for the 36 unique phenotypes
  created during the run with {\tt SlicePow}=0.96, and the HAVEGE
  entropy source.  Three organisms appear with complexity greater than
  the ancestral organism 0080aaa within the first fifty million
  timesteps, including one with a complexity nearly twice that of the
  ancestor, and one appears at $4.1\times10^9$ after which only simple
  organisms originate.}
\label{Havege-time-run}
\end{figure*}

Tierra was run with \verb+SlicePow+=0.9,0.91\ldots1.0 for $10^{10}$
timesteps (instructions executed), with a soupsize of 131072, both
with the original random number generator, and with HAVEGE. When
\verb+SlicePow+=1, the runs terminated early due to the soup dying.
Figure \ref{Havege-time-run} shows the results for {\tt SlicePow}=0.96
using the Havege generator, which produced the maximum complexity of
any run.

\begin{figure*}
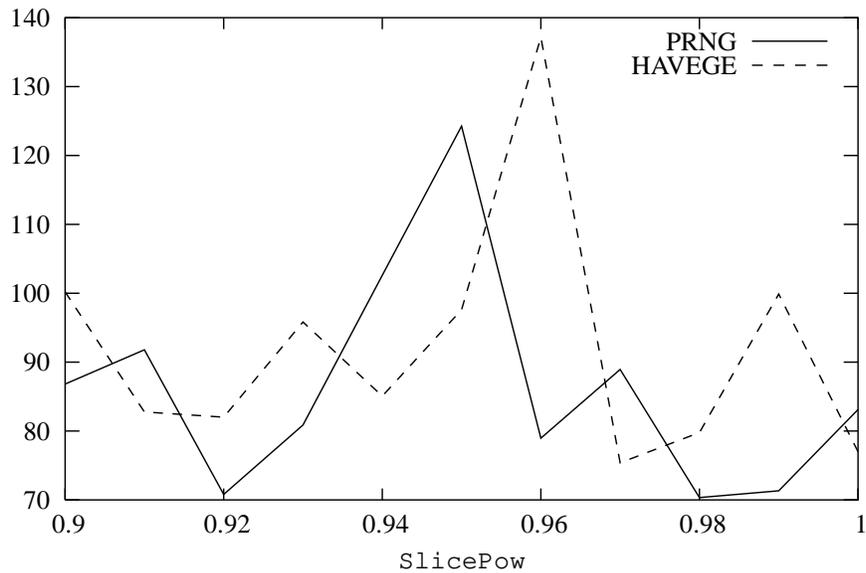

\begin{center}
% GNUPLOT: LaTeX picture using PSTRICKS macros
% Define new PST objects, if not already defined
\ifx\PSTloaded\undefined
\def\PSTloaded{t}
\psset{arrowsize=.01 3.2 1.4 .3}
\psset{dotsize=.01}
\catcode`@=11

\newpsobject{PST@Border}{psline}{linewidth=.0015,linestyle=solid}
\newpsobject{PST@Axes}{psline}{linewidth=.0015,linestyle=dotted,dotsep=.004}
\newpsobject{PST@Solid}{psline}{linewidth=.0015,linestyle=solid}
\newpsobject{PST@Dashed}{psline}{linewidth=.0015,linestyle=dashed,dash=.01 .01}
\newpsobject{PST@Dotted}{psline}{linewidth=.0025,linestyle=dotted,dotsep=.008}
\newpsobject{PST@LongDash}{psline}{linewidth=.0015,linestyle=dashed,dash=.02 .01}
\newpsobject{PST@Diamond}{psdots}{linewidth=.001,linestyle=solid,dotstyle=square,dotangle=45}
\newpsobject{PST@Filldiamond}{psdots}{linewidth=.001,linestyle=solid,dotstyle=square*,dotangle=45}
\newpsobject{PST@Cross}{psdots}{linewidth=.001,linestyle=solid,dotstyle=+,dotangle=45}
\newpsobject{PST@Plus}{psdots}{linewidth=.001,linestyle=solid,dotstyle=+}
\newpsobject{PST@Square}{psdots}{linewidth=.001,linestyle=solid,dotstyle=square}
\newpsobject{PST@Circle}{psdots}{linewidth=.001,linestyle=solid,dotstyle=o}
\newpsobject{PST@Triangle}{psdots}{linewidth=.001,linestyle=solid,dotstyle=triangle}
\newpsobject{PST@Pentagon}{psdots}{linewidth=.001,linestyle=solid,dotstyle=pentagon}
\newpsobject{PST@Fillsquare}{psdots}{linewidth=.001,linestyle=solid,dotstyle=square*}
\newpsobject{PST@Fillcircle}{psdots}{linewidth=.001,linestyle=solid,dotstyle=*}
\newpsobject{PST@Filltriangle}{psdots}{linewidth=.001,linestyle=solid,dotstyle=triangle*}
\newpsobject{PST@Fillpentagon}{psdots}{linewidth=.001,linestyle=solid,dotstyle=pentagon*}
\newpsobject{PST@Arrow}{psline}{linewidth=.001,linestyle=solid}
\catcode`@=12

\fi
\psset{unit=5.0in,xunit=5.0in,yunit=3.0in}
\pspicture(0.000000,0.000000)(1.000000,1.000000)
\ifx\nofigs\undefined
\catcode`@=11

\PST@Border(0.1170,0.1260)
(0.1320,0.1260)

\PST@Border(0.9470,0.1260)
(0.9320,0.1260)

\rput[r](0.1010,0.1260){ 70}
\PST@Border(0.1170,0.2463)
(0.1320,0.2463)

\PST@Border(0.9470,0.2463)
(0.9320,0.2463)

\rput[r](0.1010,0.2463){ 80}
\PST@Border(0.1170,0.3666)
(0.1320,0.3666)

\PST@Border(0.9470,0.3666)
(0.9320,0.3666)

\rput[r](0.1010,0.3666){ 90}
\PST@Border(0.1170,0.4869)
(0.1320,0.4869)

\PST@Border(0.9470,0.4869)
(0.9320,0.4869)

\rput[r](0.1010,0.4869){ 100}
\PST@Border(0.1170,0.6071)
(0.1320,0.6071)

\PST@Border(0.9470,0.6071)
(0.9320,0.6071)

\rput[r](0.1010,0.6071){ 110}
\PST@Border(0.1170,0.7274)
(0.1320,0.7274)

\PST@Border(0.9470,0.7274)
(0.9320,0.7274)

\rput[r](0.1010,0.7274){ 120}
\PST@Border(0.1170,0.8477)
(0.1320,0.8477)

\PST@Border(0.9470,0.8477)
(0.9320,0.8477)

\rput[r](0.1010,0.8477){ 130}
\PST@Border(0.1170,0.9680)
(0.1320,0.9680)

\PST@Border(0.9470,0.9680)
(0.9320,0.9680)

\rput[r](0.1010,0.9680){ 140}
\PST@Border(0.1170,0.1260)
(0.1170,0.1460)

\PST@Border(0.1170,0.9680)
(0.1170,0.9480)

\rput(0.1170,0.0840){ 0.9}
\PST@Border(0.2830,0.1260)
(0.2830,0.1460)

\PST@Border(0.2830,0.9680)
(0.2830,0.9480)

\rput(0.2830,0.0840){ 0.92}
\PST@Border(0.4490,0.1260)
(0.4490,0.1460)

\PST@Border(0.4490,0.9680)
(0.4490,0.9480)

\rput(0.4490,0.0840){ 0.94}
\PST@Border(0.6150,0.1260)
(0.6150,0.1460)

\PST@Border(0.6150,0.9680)
(0.6150,0.9480)

\rput(0.6150,0.0840){ 0.96}
\PST@Border(0.7810,0.1260)
(0.7810,0.1460)

\PST@Border(0.7810,0.9680)
(0.7810,0.9480)

\rput(0.7810,0.0840){ 0.98}
\PST@Border(0.9470,0.1260)
(0.9470,0.1460)

\PST@Border(0.9470,0.9680)
(0.9470,0.9480)

\rput(0.9470,0.0840){ 1}
\PST@Border(0.1170,0.1260)
(0.9470,0.1260)
(0.9470,0.9680)
(0.1170,0.9680)
(0.1170,0.1260)

\rput(0.5320,0.0210){{\tt SlicePow}}
\rput[r](0.8200,0.9270){PRNG}
\PST@Solid(0.8360,0.9270)
(0.9150,0.9270)

\PST@Solid(0.1170,0.3283)
(0.1170,0.3283)
(0.2000,0.3881)
(0.2830,0.1356)
(0.3660,0.2568)
(0.4490,0.5182)
(0.5320,0.7786)
(0.6150,0.2338)
(0.6980,0.3538)
(0.7810,0.1301)
(0.8640,0.1419)
(0.9470,0.2838)

\rput[r](0.8200,0.8850){HAVEGE}
\PST@Dashed(0.8360,0.8850)
(0.9150,0.8850)

\PST@Dashed(0.1170,0.4893)
(0.1170,0.4893)
(0.2000,0.2796)
(0.2830,0.2705)
(0.3660,0.4366)
(0.4490,0.3073)
(0.5320,0.4580)
(0.6150,0.9338)
(0.6980,0.1908)
(0.7810,0.2436)
(0.8640,0.4854)
(0.9470,0.2102)

\catcode`@=12
\fi
\endpspicture
\end{center}
\caption{Plot of the maximum $C_\mathrm{ss}$ recorded as a function of
  parsimony pressure. PRNG = pseudo
  random number generator, and HAVEGE is the entropy harvester
  mentioned in the text.}
\label{Maxcomplexity}
\end{figure*}

Figure \ref{Maxcomplexity} shows the maximum value of $C_\mathrm{ss}$
recorded for each run, as a function of parsimony pressure. It is
showing fairly clearly that a {\tt SlicePow} value between 0.95--0.96
is needed to generate additional complexity. It is also suggestive
that the entropy gatherer generates additional complexity over the
pseudo random number generator, however this effect is unlikely to be
statistically significant in this data set.

\begin{figure*}
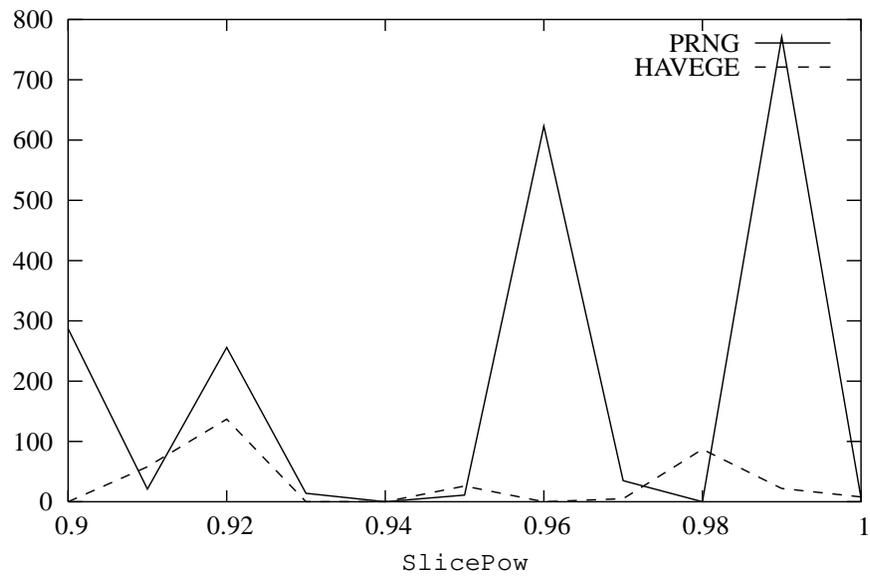

\begin{center}
% GNUPLOT: LaTeX picture using PSTRICKS macros
% Define new PST objects, if not already defined
\ifx\PSTloaded\undefined
\def\PSTloaded{t}
\psset{arrowsize=.01 3.2 1.4 .3}
\psset{dotsize=.01}
\catcode`@=11

\newpsobject{PST@Border}{psline}{linewidth=.0015,linestyle=solid}
\newpsobject{PST@Axes}{psline}{linewidth=.0015,linestyle=dotted,dotsep=.004}
\newpsobject{PST@Solid}{psline}{linewidth=.0015,linestyle=solid}
\newpsobject{PST@Dashed}{psline}{linewidth=.0015,linestyle=dashed,dash=.01 .01}
\newpsobject{PST@Dotted}{psline}{linewidth=.0025,linestyle=dotted,dotsep=.008}
\newpsobject{PST@LongDash}{psline}{linewidth=.0015,linestyle=dashed,dash=.02 .01}
\newpsobject{PST@Diamond}{psdots}{linewidth=.001,linestyle=solid,dotstyle=square,dotangle=45}
\newpsobject{PST@Filldiamond}{psdots}{linewidth=.001,linestyle=solid,dotstyle=square*,dotangle=45}
\newpsobject{PST@Cross}{psdots}{linewidth=.001,linestyle=solid,dotstyle=+,dotangle=45}
\newpsobject{PST@Plus}{psdots}{linewidth=.001,linestyle=solid,dotstyle=+}
\newpsobject{PST@Square}{psdots}{linewidth=.001,linestyle=solid,dotstyle=square}
\newpsobject{PST@Circle}{psdots}{linewidth=.001,linestyle=solid,dotstyle=o}
\newpsobject{PST@Triangle}{psdots}{linewidth=.001,linestyle=solid,dotstyle=triangle}
\newpsobject{PST@Pentagon}{psdots}{linewidth=.001,linestyle=solid,dotstyle=pentagon}
\newpsobject{PST@Fillsquare}{psdots}{linewidth=.001,linestyle=solid,dotstyle=square*}
\newpsobject{PST@Fillcircle}{psdots}{linewidth=.001,linestyle=solid,dotstyle=*}
\newpsobject{PST@Filltriangle}{psdots}{linewidth=.001,linestyle=solid,dotstyle=triangle*}
\newpsobject{PST@Fillpentagon}{psdots}{linewidth=.001,linestyle=solid,dotstyle=pentagon*}
\newpsobject{PST@Arrow}{psline}{linewidth=.001,linestyle=solid}
\catcode`@=12

\fi
\psset{unit=5.0in,xunit=5.0in,yunit=3.0in}
\pspicture(0.000000,0.000000)(1.000000,1.000000)
\ifx\nofigs\undefined
\catcode`@=11

\PST@Border(0.1170,0.1260)
(0.1320,0.1260)

\PST@Border(0.9470,0.1260)
(0.9320,0.1260)

\rput[r](0.1010,0.1260){ 0}
\PST@Border(0.1170,0.2313)
(0.1320,0.2313)

\PST@Border(0.9470,0.2313)
(0.9320,0.2313)

\rput[r](0.1010,0.2313){ 100}
\PST@Border(0.1170,0.3365)
(0.1320,0.3365)

\PST@Border(0.9470,0.3365)
(0.9320,0.3365)

\rput[r](0.1010,0.3365){ 200}
\PST@Border(0.1170,0.4418)
(0.1320,0.4418)

\PST@Border(0.9470,0.4418)
(0.9320,0.4418)

\rput[r](0.1010,0.4418){ 300}
\PST@Border(0.1170,0.5470)
(0.1320,0.5470)

\PST@Border(0.9470,0.5470)
(0.9320,0.5470)

\rput[r](0.1010,0.5470){ 400}
\PST@Border(0.1170,0.6523)
(0.1320,0.6523)

\PST@Border(0.9470,0.6523)
(0.9320,0.6523)

\rput[r](0.1010,0.6523){ 500}
\PST@Border(0.1170,0.7575)
(0.1320,0.7575)

\PST@Border(0.9470,0.7575)
(0.9320,0.7575)

\rput[r](0.1010,0.7575){ 600}
\PST@Border(0.1170,0.8628)
(0.1320,0.8628)

\PST@Border(0.9470,0.8628)
(0.9320,0.8628)

\rput[r](0.1010,0.8628){ 700}
\PST@Border(0.1170,0.9680)
(0.1320,0.9680)

\PST@Border(0.9470,0.9680)
(0.9320,0.9680)

\rput[r](0.1010,0.9680){ 800}
\PST@Border(0.1170,0.1260)
(0.1170,0.1460)

\PST@Border(0.1170,0.9680)
(0.1170,0.9480)

\rput(0.1170,0.0840){ 0.9}
\PST@Border(0.2830,0.1260)
(0.2830,0.1460)

\PST@Border(0.2830,0.9680)
(0.2830,0.9480)

\rput(0.2830,0.0840){ 0.92}
\PST@Border(0.4490,0.1260)
(0.4490,0.1460)

\PST@Border(0.4490,0.9680)
(0.4490,0.9480)

\rput(0.4490,0.0840){ 0.94}
\PST@Border(0.6150,0.1260)
(0.6150,0.1460)

\PST@Border(0.6150,0.9680)
(0.6150,0.9480)

\rput(0.6150,0.0840){ 0.96}
\PST@Border(0.7810,0.1260)
(0.7810,0.1460)

\PST@Border(0.7810,0.9680)
(0.7810,0.9480)

\rput(0.7810,0.0840){ 0.98}
\PST@Border(0.9470,0.1260)
(0.9470,0.1460)

\PST@Border(0.9470,0.9680)
(0.9470,0.9480)

\rput(0.9470,0.0840){ 1}
\PST@Border(0.1170,0.1260)
(0.9470,0.1260)
(0.9470,0.9680)
(0.1170,0.9680)
(0.1170,0.1260)

\rput(0.5320,0.0210){{\tt SlicePow}}
\rput[r](0.8200,0.9270){PRNG}
\PST@Solid(0.8360,0.9270)
(0.9150,0.9270)

\PST@Solid(0.1170,0.4281)
(0.1170,0.4281)
(0.2000,0.1481)
(0.2830,0.3954)
(0.3660,0.1407)
(0.4490,0.1260)
(0.5320,0.1376)
(0.6150,0.7817)
(0.6980,0.1628)
(0.7810,0.1260)
(0.8640,0.9375)
(0.9470,0.1323)

\rput[r](0.8200,0.8850){HAVEGE}
\PST@Dashed(0.8360,0.8850)
(0.9150,0.8850)

\PST@Dashed(0.1170,0.1260)
(0.1170,0.1260)
(0.2000,0.1870)
(0.2830,0.2702)
(0.3660,0.1260)
(0.4490,0.1260)
(0.5320,0.1534)
(0.6150,0.1260)
(0.6980,0.1313)
(0.7810,0.2176)
(0.8640,0.1492)
(0.9470,0.1344)

\catcode`@=12
\fi
\endpspicture
\end{center}
\caption{Plot of the origination time of the organism with maximal
  complexity, as a function of parsimony pressure. PRNG = pseudo
  random number generator, and HAVEGE is the entropy harvester
  mentioned in the text.}
\label{origtime}
\end{figure*}

A different view of the data can be seen in figure \ref{origtime},
where the origination of the organism of maximal complexity is
plotted. The interesting thing here is that the pseudo random number
generator took a lot longer to find its maximally complex organism
than the entropy gatherer algorithm.

\section{Discussion}

The results reported here are of a small scale pilot study
to study the effect of parsimony and of random oracles. Obviously much
work needs to be done to tighten up the methodology of the experiment,
and to perform analysis of statistical significance on the
results. Clearly, this experiment does not show evidence of open-ended
complexity growth either. Nevertheless, these interim results look encouraging.

Since the Tierra simulations are run in parallel on a Linux cluster,
and it doesn't matter if one uses the same sequence of random numbers
in each simulation, it should be possible to combine the entropy
harvested from all CPUs, as well as network latency on the
interconnect to substantially increase the entropy production of
HAVEGE. This will require substantial recoding of HAVEGE to exploit this. 

In the history of Earth's biosphere, complexity of individuals remain
largely static for the first 2 billion years of life. It was only with
originations of Eukaryotes circa 2Gya and and of multicellular life
circa 600Mya that we have any appreciable jump in complexity over the
previous 2 billion years of bacterial life. During Phanerozoic
(540Mya--present),  there is little unambiguous evidence of complexity
growth of organisms\cite{McShea96}. However, what is a very clear
trend is growth in the complexity of ecosystems. Diversity of the
Earth's biosphere appears to have grown exponentially since Cambrian
times \cite{Benton01}. 

Looking at things another way, a multicellular animal can be
considered as an ecosystem of eukaryotic cells (OK so the genetic code
for most of the cells is identical --- gut flora being the obvious
exception), and each eukaryotic cell can be considered an ecosystem of
bacterial cells (nucleus + organelles). If anything, the ``parts'' of
the World's biosphere have gotten simpler --- it's the network
connecting the parts that shown the complexity growth. In the Tierra
case, what we should be looking for is overall complexity of the
ecosystem, not complexity of the individual digital organisms.

At present, we still don't have a good theory for how to measure the
complexity of an ecosystem, knowing its foodweb. Diversity is a lower
bound on complexity, but is a rather crude indicator of overall
complexity. Bedau-Packard \cite{Bedau-etal98} statistics use diversity
as one of the key indicators of open-ended evolution. This is probably
all that is ever likely to be available for the Earth's biosphere, but
in artificial life systems we can look for better measures of
ecosystem complexity.

\section*{Acknowledgment}

I would like to thank the {\em Australian Centre for Advanced
  Computing and Communications} for a grant of time on their Linux
cluster for performing this work.

\bibliographystyle{alife9}
\bibliography{rus}
\end{document}